\documentclass[12pt]{iopart}
\usepackage{setstack}
\usepackage{iopams}
\usepackage{amssymb}
\usepackage{graphicx}
\begin{document}
%--------------------------------------------------------------------
% Local definitions
\def\PRL#1{{\it Phys.\ Rev.\ Lett.}\ {\bf#1}}
\def\JCAP#1#2{#2 {\it J.\ Cosmol.\ Astropart.\ Phys.}\ JCAP\,{#1}\,(#2)\,}
\def\ApJ#1{{\it Astrophys.\ J.}\ {\bf#1}}
\def\PR#1#2{{\it Phys.\ Rev.}\ #1~{\bf#2}}
\def\MNRAS#1{{\it Mon.\ Not.\ R.\ Astr.\ Soc.}\ {\bf#1}}
\def\CQG#1{{\it Class.\ Quantum Grav.}\ {\bf#1}}
\def\GRG#1{{\it Gen.\ Relativ.\ Grav.}\ {\bf#1}}
\def\IJMP#1{{\it Int.\ J.\ Mod.\ Phys.}\ {\bf#1}}
\def\beq{\begin{equation}} \def\eeq{\end{equation}}
\def\bea{\begin{eqnarray}} \def\eea{\end{eqnarray}}
\def\Z#1{_{\lower2pt\hbox{$\scriptstyle#1$}}} \def\w#1{\,\hbox{#1}}
\def\X#1{_{\lower2pt\hbox{$\scriptscriptstyle#1$}}}
\font\sevenrm=cmr7 \def\ns#1{_{\hbox{\sevenrm #1}}}
\def\Ns#1{\Z{\hbox{\sevenrm #1}}} \def\ave#1{\langle{#1}\rangle}
\def\lsim{\mathop{\hbox{${\lower3.8pt\hbox{$<$}}\atop{\raise0.2pt\hbox{$\sim$}}
$}}} \def\gsim{\mathop{\hbox{${\lower3.8pt\hbox{$>$}}\atop{\raise0.2pt\hbox{$
\sim$}}$}}} \def\kms{\w{km}\;\w{sec}^{-1}}\def\kmsMpc{\kms\w{Mpc}^{-1}}
\def\dd{{\rm d}} \def\ds{\dd s} \def\etal{{\em et al}}
\def\al{\alpha}\def\be{\beta}\def\ga{\gamma}\def\de{\delta}\def\ep{\epsilon}
\def\et{\eta}\def\th{\theta}\def\ph{\phi}\def\rh{\rho}\def\si{\sigma}
\def\ta{\tau} \def\tn{\ts\Z0}
\def\frn#1#2{{\textstyle{#1\over#2}}} \def\Deriv#1#2#3{{#1#3\over#1#2}}
\def\Der#1#2{{#1\hphantom{#2}\over#1#2}} \def\pt{\partial} \def\ab{{\bar a}}
\def\goesas{\mathop{\sim}\limits} \def\tw{\ta\ns{w}}
\def\gb{\bar\ga}\def\omi{\OM_i} \def\I{{\hbox{$\scriptscriptstyle I$}}}
\def\av{{a\ns{v}\hskip-2pt}} \def\aw{{a\ns{w}\hskip-2.4pt}}\def\Vav{{\cal V}}
\def\DD{{\cal D}}\def\gd{{{}^3\!g}}\def\half{\frn12}\def\Rav{\ave{\cal R}}
\def\QQ{{\cal Q}}\def\dsp{\displaystyle} \def\rw{r\ns w}
\def\mean#1{{\vphantom{\tilde#1}\bar#1}}\def\bx{{\mathbf x}}\def\mx{\mean x}
\def\bn{\mean n}\def\bH{\mean H}\def\Hb{\bH\Z{\!0}}\def\gb{\mean\ga}
\def\bD{\mean D}\def\rhb{\mean\rh}\def\OM{\mean\Omega}
\def\fw{{f\ns w}}\def\fv{{f\ns v}} \def\goesas{\mathop{\sim}\limits}
\def\fvn{f\ns{v0}} \def\fvf{\left(1-\fv\right)}
\def\OMM{\OM\Z M}\def\OMk{\OM\Z k}\def\OMQ{\OM\Z{\QQ}}
\def\OmMn{\Omega_{\ns{M0}}}\def\OMBn{\OM_{\ns{B0}}}\def\OMCn{\OM_{\ns{C0}}}
\def\OMMn{\OM_{\ns{M0}}}\def\OMkn{\OM_{\ns{k0}}}\def\OMn{\OM_{\ns{0}}}
\def\OmBn{\Omega_{\ns{B0}}}\def\OmCn{\Omega_{\ns{C0}}}\def\OMRn{\OM_{\ns{R0}}}
\def\ts{t} \def\tb{\ts'}\def\tc{\tau}\def\la{\lambda}
\def\Hm{H\Z0}\def\fvi{{f\ns{vi}}}\def\fwi{{f\ns{wi}}}
\def\Hv{H\ns v}\def\Hw{H\ns w}\def\kv{k\ns v}\def\hri{h_{ri}}
\def\LA{\Lambda}\def\LCDM{$\LA$CDM}\def\Aa{{\cal A}}
\def\zb{\bar{z}}\def\nb{\bar{n}}\def\Tb{\bar{T}}
\def\rhM{\rh_{\ns M}}\def\rhbMn{\rhb_{\ns{M0}}}\def\rhR{\rh_{\ns R}}
\def\rhbRn{\rhb_{\ns{R0}}} \def\pR{P_{\ns R}} \def\bal{\bar\al}
\def\OMRn{\OM\Z{R0}}\def\OMR{\OM\Z R}\def\abn{\ab_{\ns0}}\def\gbn{\gb_{\ns0}}
\def\Fx{F}\def\fFx{1+x^2{\Fx}^2}\def\Oma{\al_{\ns{M0}}}
\def\Ora{\al_{\ns{R0}}} \def\kB{k\Z B}\def\Fn{\Fx_{\ns0}}\def\xn{x_{\ns0}}
\def\h{\,h^{-1}}\def\hm{\h\hbox{Mpc}} \def\nbB{\nb_{\ns{B}}}
\def\etBg{\et\Z{B\ga}}\def\GA{\Gamma}\def\tao{\tau_{\ns o}}\def\mpr{m\Ns p}
\def\OMPn{\OM\Z{\ga0}}\def\tad{\tau_{\ns d}}\def\vp{\vphantom{\Bigl|}}
%----------------------------------------------------------------
\title[Timescape cosmology with radiation fluid]{Timescape cosmology with
radiation fluid}
%-----------------------------------------------------------------
\author{James A.G.~Duley, M.~Ahsan Nazer and David L. Wiltshire}
%-----------------------------------------------------------------
\address{Department of Physics and Astronomy, University of Canterbury,
Private Bag 4800, Christchurch 8140, New Zealand}
%-----------------------------------------------------------------
%\eads{\mailto{David.Wiltshire@canterbury.ac.nz}\\
%http://www2.phys.canterbury.ac.nz/$\goesas$dlw24/}

\begin{abstract}
The timescape cosmology represents a potentially viable alternative to
the standard homogeneous cosmology, without the need for dark energy.
Although average cosmic evolution in the timescape scenario only differs
substantially from that of Friedmann-Lema\^{\i}tre model at relatively
late epochs when the contribution from the energy density of radiation
is negligible, a full solution of the Buchert equations to incorporate
radiation is necessary to smoothly match parameters to the epoch of
photon decoupling and to obtain constraints from cosmic microwave background
data. Here we extend the matter-dominated solution found in earlier work to
include radiation, providing series solutions at early times and an efficient
numerical integration strategy for generating the complete solution. The
numerical solution is used to directly calculate the scale of the sound
horizon at decoupling, and at the baryon drag epoch. The constraints on
these scales from the Planck satellite data yield bounds on the timescape
cosmological parameters, which are found to also agree with the best-fit
values from a recent analysis of SDSS-II supernova data, while avoiding the
problem of a primordial lithium-7 abundance anomaly.
\end{abstract}
\bigskip

%Keywords: Cosmology, backreaction, averaging, dark energy
%-----------------------------------------------------------------
\maketitle
%-----------------------------------------------------------------

\section{Introduction}

The standard model of cosmology is based on the assumption that average
cosmic evolution is identical to that of an exactly homogeneous isotropic
Friedmann--Lema\^{\i}tre--Robertson--Walker (FLRW) model. This assumption
is well justified at the epoch of last scattering, by the evidence of the
Cosmic Microwave Background (CMB) radiation. However, small density
perturbations eventually grow nonlinear, forming the observed structures
of the universe, and by the present epoch the universe is only homogeneous
in some average statistical sense when one averages on scales $\gsim100\hm$,
where $h$ is the dimensionless parameter related to the Hubble constant by
$\Hm=100h\kmsMpc$. Below this scale we observe a universe dominated in
volume by voids \cite{HV1}--\cite{Pan11}, with clusters of galaxies in walls,
sheets and filaments surrounding and threading the voids.

The problem of fitting a smooth geometry to this complex hierarchical
structure entails many fundamental issues \cite{fit1}--\cite{dust}, including
in particular: (i) how is average cosmic evolution to be described; and
(ii) how are local observables related to quantities defined with respect
to some average geometry? There has been considerable interest in these
problems in recent years (for some recent reviews see \cite{dust}--\cite{BR})
since it is possible that a full understanding of these issues might explain
the observation of cosmic acceleration attributed to a smooth form of dark
energy in the standard cosmology.

In this paper, we will focus on the timescape cosmology \cite{clocks}--\cite
{obs}, which is a phenomenologically viable model of the universe to the
extent that it has been tested \cite{obs}--\cite{grb}. The differences
between the predictions of the timescape model and those of the \LCDM\
model for supernova luminosity distances are at the same level as current
systematic uncertainties in data reduction \cite{SW}. The timescape model
fits the angular scale of sound horizon in the CMB anisotropy data, and
the Baryon Acoustic Oscillation (BAO) scale in galaxy clustering statistics
\cite{LNW} but these tests have not yet been developed to the extent
that they can tightly constrain cosmological parameters.

The timescape model is based
on the Buchert scheme \cite{buch1,buch2} for statistical averages of a fully
inhomogeneous geometry, while maintaining a statistical Copernican principle.
Since the Buchert scheme involves statistical quantities, additional physical
assumptions are required to relate its average parameters to cosmological
observables \cite{BC1,BC2}.

In the timescape model it is postulated that the relevant physical assumptions
relate to gravitational energy: in particular, to the relative regional volume
deceleration of expanding regions of different density,
which provides a measure of the relative kinetic energy of regional expansion
\cite{clocks,equiv}. In the absence of an exact timelike
Killing vector, bound systems -- which necessarily form in regions where the
density is greater than critical -- can always be embedded within expanding
regions bounded by a ``finite infinity surface'' \cite{fit1,clocks,equiv}
within which the smoothed geometry is spatially flat, with a close to
Einstein-de Sitter expansion law.

It is postulated that in describing the statistical cosmological geometry
one can always choose a uniform Hubble flow slicing, akin to a constant
mean extrinsic curvature (CMC) slicing, in which the effects of regional
scalar spatial curvature are compensated by the choice of the canonical time
coordinate of ``cosmological inertial frames'', namely expanding regions whose
spatial extent is smaller than the (negative) curvature scale but larger than
bound systems \cite{equiv}. The canonical time parameter for observers within
finite infinity regions, where galaxies and other bound systems are located,
is then related to the time parameter appearing in the Buchert statistical
averages by a phenomenological lapse function. By a procedure of matching
null geodesics in the two geometries \cite{clocks}, solutions of the
Buchert equations can be related to cosmological observables determined by
observers (such as ourselves) within bound systems where the regional spatial
curvature is different from the global statistical average.

The physical explanation of apparent cosmic acceleration in the timescape
scenario relies then not simply on the backreaction of inhomogeneities which
define the average cosmic evolution, but more on the differences of
gravitational energy manifest in the canonical clocks of observers in galaxies
as compared to observers in voids, where the spatial curvature is negative.
These differences are insignificant in the early universe which is close to
homogeneous, but the differences grow cumulatively and become especially large
when voids come to dominate the volume of the universe. Phenomenologically,
apparent acceleration is found to begin when the void fraction reaches 59\%
\cite{clocks}.

The timescape scenario faces two main challenges to be developed into a model
that can fully compete with the standard \LCDM\ cosmology:\begin{itemize}
\item At a formal level new mathematical constructions are required to
define a modified statistical geometry of the universe, and the methods by
which it is patched to regional geometries. The procedures of
coarse-graining, and their relationship to gravitational energy and entropy
have to be well-understood.
\item At an observational level, cosmological tests which rely heavily on the
standard FLRW model in data reduction procedures need to be revisited
from first principles. This applies in particular to the analysis of the
power spectrum of CMB anisotropies, and to the analysis of galaxy clustering
statistics.
\end{itemize}
The present paper will take steps towards the second of these goals by fully
incorporating a radiation fluid in the solution of the Buchert equations.

At epochs prior to last scattering the universe is close to homogeneous, so
that timescape model is almost indistinguishable from the standard cosmology.
In previous work \cite{clocks} estimates of the angular diameter distance of
the sound horizon, and calibrations of the baryon--to--photon ratio for big
bang nucleosynthesis, were made by simply matching the matter only solution
of the Buchert equations to a spatially flat FLRW model with matter and
radiation. While this may be sufficient for simple estimates, a more detailed
treatment of cosmic evolution of the early universe after last scattering
requires that the radiation component is incorporated directly.

\section{Buchert equations for two-scale model with radiation fluid}
Our primary aim in this paper is to extend the exact solution of
\cite{sol,obs} to include the contributions of relativistic species
(photons and neutrinos) directly in the solution of the Buchert equations.
The matter content will therefore be taken as that of the standard
cosmology without a cosmological constant, namely matter fields in the
form of both baryonic and nonbaryonic matter treated as dust, plus photons
and the standard three generations of neutrinos.

At early epochs when radiation is dominant the universe is assumed to be very
close to homogeneous and isotropic, and thus the solution we expect will be
very close to that of a standard matter plus radiation FLRW model with
negligible spatial curvature. Furthermore at late epochs, when the solutions to
the Buchert equations differ substantially from a FLRW model, the contribution
of the radiation energy density to the overall energy density is negligible.
At late epochs it is only the matter component which drives the overall
evolution of the universe (assuming no dark energy), and it is the matter
component which defines the density gradients. While the radiation fluid
certainly responds to density gradients, this only affects
questions such as gravitational lensing, rather than the average
cosmological evolution considered here.

We will therefore treat the radiation fluid as a component with a pressure $\pR
=\frac13\rhR$ which commutes under the Buchert average,
\beq
\pt_t\ave{\pR}-\ave{\pt_t \pR}=\ave{\pR\th}-\ave{\pR}\ave{\th}=0,
\eeq
throughout the evolution of the universe, rather than using the more detailed
Buchert formalism that applies to fluids with pressure \cite{buch2}. Here $\th$
is the expansion scalar and angle brackets denote the spatial volume average
of a quantity on the surface of average homogeneity, so that $\ave{\pR}\equiv
\left(\int_\DD\dd^3x\sqrt{\det\gd}\,{\pR}(t,\bx)\right)/\Vav(t)$, where $\Vav
(t)\equiv\int_\DD\dd^3x\sqrt{\det\gd}$ is the average spatial volume,
$\gd_{ij}$ being the 3-metric. The detailed Buchert
formalism for general averaging and backreaction in fluids with pressure may
be of relevance for deriving further results in a perturbation theory approach
in the early universe. In this paper, however, we confine ourselves to
finding a smooth solution which makes a transition from radiation domination
to the late epoch matter-dominated solution of the timescape cosmology
\cite{clocks}--\cite{obs}.

With our assumptions the radiation fluid does not contribute to the
backreaction, and the Buchert equations \cite{buch1,buch2} may then be
written
\bea
&&\dsp{3\dot\ab^2\over\ab^2}=8\pi G\left(\ave{\rhM}+\ave{\rhR}\right)-\half
\Rav-\half\QQ,\label{b1}\\
&&{3\ddot\ab\over\ab}=-4\pi G\left(\ave{\rhM}+2\ave{\rhR}\right)+\QQ,
\label{b2}\\
&&\pt_t\ave{\rhM}+3{\dot\ab\over\ab}\ave{\rhM}=0,\label{b3}\\
&&\pt_t\ave{\rhR}+4{\dot\ab\over\ab}\ave{\rhR}=0,\label{b4}\\
&&\pt_t\left(\ab^6\QQ\right)+\ab^4 \pt_t\left(\ab^2\Rav\right)=0,
\label{b5}\eea
where an overdot denotes a time derivative for volume-average observers
``comoving'' with the dust of density $\rhM$. Here $\ab(t)\equiv\left[\Vav(t)/
\Vav(t\Z0)\right]^{1/3}$ is the volume-average scale factor, $\Rav$ is the
average spatial curvature scalar and
\beq\QQ=\frn23\left(\langle\th^2\rangle-\langle\th\rangle^2\right)-
2\langle\si^2\rangle,\label{backr}\eeq
is the kinematic backreaction, which combines the variance in volume expansion
and the shear scalar $\si^2=\frn12\si_{\al\be}\si^{\al\be}$. We use units in
which $c=1$. Equation (\ref{b5}) is a condition needed to ensure that
(\ref{b1}) is the integral of (\ref{b2}). The integrability condition
(\ref{b5}) constrains just one of the two unknowns $\QQ$ and $\Rav$ in general.
In the timescape model an ensemble of wall and void regions is further
specified, thereby constraining $\Rav$ and giving a coupled set of differential
equations which can be solved.

The notion of ``comoving with the dust'' is
reinterpreted in the timescape approach. Since particle geodesics cross
during structure formation one must necessarily coarse-grain over scales
larger than galaxies to define ``dust'' in cosmology. However,
galaxies are not isolated particles whose masses remain invariant from
last-scattering until today. In the timescape approach it
is assumed that ``dust'' can only be defined as expanding fluid cells
coarse-grained at a scale a few times larger than that of the largest typical
nonlinear structures, so that the mass contained in a dust cell
does not change on average. Given that the largest typical nonlinear structures
are voids of diameter $30\hm$ \cite{HV1,HV2}, we take the coarse-graining scale
or {\em statistical homogeneity scale} to be comparable to the BAO scale,
$100\hm$. The Buchert time parameter is therefore regarded as a collective
coordinate of such a coarse-grained ``dust'' cell.
Equations (\ref{b1})--(\ref{b5}), which involve derivatives
with respect to Buchert time, represent the evolution of
a statistical geometry. The Buchert time parameter would
only be directly measured by a {\em volume-average} isotropic
observer, namely an observer who measures an isotropic CMB and
whose local regional spatial curvature scalar happens to match
the scalar curvature averaged over a horizon volume, $\Rav$.

Following references \cite{clocks,sol} we assume that the present epoch horizon
volume, $\Vav=\Vav\ns i\ab^3$, is a disjoint union of void and wall regions
characterized by scale factors $\av$ and $\aw$ related to the volume-average
scale factor by
\beq
\ab^3 = \fvi \av^3 + \fwi \aw^3 \label{bav}
\eeq
where $\fvi$ and $\fwi = 1-\fvi$ represent the fraction of the initial volume,
$\Vav\ns i$, in void and wall regions respectively at an early unspecified
epoch. The voids are assumed to have negative spatial curvature characterized
by $\Rav_{\ns v}\equiv 6\kv/\av^2$ with $\kv<0$, while the wall regions
\cite{clocks} are on average spatially flat, $\Rav_{\ns w}=0$.

In previous work the initial volume was assumed to be prescribed at the surface
of last scattering. Furthermore, since finite infinity regions are only
well--defined once gravitational collapse results in the formation of bound
structures, the operational definition of $\fwi$ and $\fvi$ is complex.
Following \cite{clocks}--\cite{obs} we assume that $\fwi$ is close to unity,
consistent with the universe at last scattering being very close to a spatially
flat FLRW model. The tiny void fraction $\fvi\ll 1$ then represents that
fraction of the present epoch horizon volume in which underdense perturbations
were not compensated by overdense perturbations at last scattering. It is
convenient to rewrite (\ref{bav}) as
\beq
\fv(t)+\fw(t) = 1,
\eeq
where $\fw(t)=\fwi \aw^3/\ab^3$ is the wall volume fraction and $\fv(t)=\fvi\aw
^3/\ab^3$ is the void volume fraction. Since $\ave{\rhM}=\rhbMn(\ab/\abn)^{-3}$
and $\ave{\rhR}=\rhbRn(\ab/\abn)^{-4}$, where the subscript zero refers to
quantities evaluated at the present epoch, after solving (\ref{b3}) and
(\ref{b4}) in the standard fashion the remaining independent Buchert equations
in (\ref{b1})--(\ref{b5}) may be written as
\beq
\frac{\dot\ab^2}{\ab^2}+\frac{\dot\fv^2}{9\fv(1-\fv)}-\frac{\al^2\fv^{1/3}}
{\ab^2} = \frac{8\pi G}{3}\left( \rhbMn\frac{\abn^3}{\ab^3}+\rhbRn
\frac{\abn^4}{\ab^4}\right),\label{deScale}
\eeq
\beq
\ddot\fv+\frac{\dot\fv^2(2\fv-1)}{2\fv(1-\fv)} + 3\frac{\dot\ab}{\ab}\dot\fv-
\frac{3\al^2 \fv^{1/3}(1-\fv)}{2\ab^2}=0,\label{defv}
\eeq
where $\al^2\equiv-\kv\fvi^{2/3}>0$. We note that (\ref{defv}) is unchanged
from the corresponding equation in the matter only case \cite{clocks,sol}. The
acceleration equation (\ref{b2}) which may be derived from (\ref{deScale}) and
(\ref{defv}) is given by
\beq
\frac{\ddot\ab}{\ab}=\frac{2\dot\fv^2}{9\fv(1-\fv)}- \frac{4\pi G}{3}
\frac{\abn^3}{\ab^3}\left[\rhbMn+2\rhbRn\frac{\abn}{\ab}\right],
\eeq
where have made the substitutions \cite{clocks}
\beq
\Rav\equiv\frac{6\kv\fvi^{2/3}\fv^{1/3}}{\ab^2}\;,\;\;
\QQ\equiv\frac{2\dot\fv^2}{3\fv(1-\fv)}\,.\label{defs}
\eeq

The first Buchert equation (\ref{deScale}) is the equivalent of the
Friedmann equation for the bare Hubble parameter, which from (\ref{bav})
is given by
\beq
\bH\equiv{\dot\ab\over\ab}=\fw\Hw+\fv\Hv\,,\label{bareH}
\eeq
where $\Hw\equiv\dot\aw/\aw$ and $\Hv\equiv\dot\av/\av$ are the Hubble
parameters of the walls and voids respectively as determined by the clocks
of volume--average observers. These satisfy an inequality
\beq h_r\equiv\Hw/\Hv<1\,. \eeq
It is an assumption of the timescape model that while any observer with
a single clock will always determine wall regions to be expanding
at a slower rate than void regions, the actual expansion rates depend on time
parameters which differ on account of gravitational energy gradients between
regions of different spatial curvature. In particular, observers in the
denser (spatially flat) wall regions use the wall time parameter $\dd\tw=\dd t
/\gb$, where
\beq
\gb=1+\left(1-h_r\over h_r\right)\fv
\eeq
is the {\em phenomenological lapse function} relating the clock of the
wall observer to that of a volume--average observer.

Equation (\ref{deScale}) is also conveniently written in the form
\beq
\OMM+\OMR+\OMk+\OMQ=1,\label{beq1}
\eeq
where
\bea\OMM&=&{8\pi G\rhb\Z{M0}\abn^3\over 3\bH^2\ab^3}\,,\label{om1}\\
\OMR&=&{8\pi G\rhb\Z{R0}\abn^4\over 3\bH^2\ab^4}\,,\label{om2}\\
\OMk&=&{\al^2\fv^{1/3}\over \ab^2\bH^2}\,,\label{om3}\\
\OMQ&=&{-\dot\fv^2\over 9\fv(1-\fv)\bH^2}={-(1-\fv)(1-\gb)^2\over\fv\gb^2}\,,
\label{om4}
\eea
are the volume--average or ``bare'' density parameters of matter, radiation,
average spatial curvature and kinematic backreaction respectively. It would
be straightforward to add a cosmological constant term to the right hand side
of (\ref{deScale}) with the addition of a further density parameter $\OM\Z\LA=
\LA/(3\bH^2)$, and in fact the equivalent solution with matter and a
cosmological constant (but no radiation) has been derived in \cite{V12,VM}.
Since we are investigating the possibility of a viable cosmology without dark
energy, we set $\OM\Z\LA=0$.

If we evaluate (\ref{beq1})--(\ref{om4}) at the present epoch we find that the
present value of the phenomenological lapse parameter is given in terms
of the other parameters by
\beq
\gbn={\sqrt{1-\fvn}\,\left(\sqrt{1-\fvn}+\sqrt{\fvn(1-\fvn)(\OMn-1)}\right)
\over1-\fvn\OMn}\,,
\label{gbn1}\eeq
where
\beq \OMn\equiv\OMMn+\OMRn+\OMkn\,, \label{gbn2}\eeq
which satisfies $1<\OMn<\fvn^{-1}$. If we drop the subscript zero in
(\ref{gbn1}) and (\ref{gbn2}) we obtain a generic relation for $\gb$ in
terms of $\OMM$, $\OMR$ and $\OMk$ at any epoch.

The bare cosmological parameters are not those determined by observers
in galaxies in wall regions where the local spatially flat curvature is
different to the volume--average one. Instead, using a matching procedure
\cite{clocks}, the relevant dressed Hubble parameter is determined
to be
\beq H=\gb\bH-\gb^{-1}\dot\gb\,.\label{Hd}\eeq
The dressed matter density parameter $\Omega_{\ns M}\equiv
\gb^3\OMM$ takes numerical values closer to the corresponding parameter
for the concordance \LCDM\ model when evaluated at the present epoch.
\section{Solution for the two-scale model with radiation}

In the case of purely nonrelativistic matter, $\rh\Z R=0$, an analytic
solution of the ODEs (\ref{deScale}), (\ref{defv}) is readily found
\cite{sol,obs}. This is in fact a consequence of the energy density
scaling as a simple power of the total volume, as do the wall and void
fractions. In the present case, which is not so simple, no further
analytic integrals of the ODEs are easily obtained.

\subsection{Early time series solution}
Although an exact analytic solution to (\ref{deScale}) and (\ref{defv}) is not
to be found, a series solution in powers of $(\Hb t)^{1/2}$ is readily obtained
in the early time limit $t\to0$. Here $\Hb\equiv\bH(\tn)$ is the bare Hubble
constant. We find
\bea
% \bar{a} series
\fl\frac\ab{\abn}=&\sqrt2\,\OMRn^{1/4}(\Hb t)^{1/2}+\frac{\OMMn(\Hb t)}{3\,
\OMRn^{1/2}}-\frac{7\OMMn^2(\Hb t)^{3/2}}{72\sqrt2\,\OMRn^{5/4}}
\nonumber\\\fl&
+\left(\frac{5\,\OMMn^3}{216\,\OMRn^{3/2}}+\frac{8\bal^3}{5\sqrt5}
\right)\frac{(\Hb t)^2}{\OMRn^{1/2}}
-\left(\frac{91\,\OMMn^3}{6912\sqrt2\,\OMRn^{3/2}}+\frac{7
\sqrt2\,\bal^3}{75\sqrt5}\right)\frac{\OMMn(\Hb t)^{5/2}}{\OMRn^{5/4}}
\nonumber\\\fl&\quad
+\left(\frac{\OMMn^3}{243\,\OMRn^{3/2}}+\frac{604\,\bal^3}{8925
\sqrt5}\right)\frac{\OMMn^2(\Hb t)^3}{\OMRn^2}+\cdots\label{absol}\\
% fv series
\fl\frac\fv{\bal^3}=&\frac{2\sqrt2(\Hb t)^{3/2}}{5\sqrt5\,\OMRn^{3/4}}-\frac{8
\,\OMMn(\Hb t)^2}{25\sqrt5\,\OMRn^{3/2}}\nonumber\\\fl&
+\frac{6617\,\OMMn^2(\Hb t)^{5/2}}{25500\sqrt{10}\,\OMRn^{9/4}}
-\left(\frac{127\sqrt5\,\OMMn^3}{5967\,\OMRn^{3/2}}+\frac{576\,\bal^3}
{8125}\right)\frac{(\Hb t)^3}{\OMRn^{3/2}}\nonumber\\\fl&\quad
+\left(\frac{8811748927\,\OMMn^3}{100086480000\sqrt{10}\,\OMRn^{3/2}} +
\frac{88522\sqrt2\,\bal^3}{1503125}\right)\frac{\OMMn(\Hb t)^{7/2}}
{\OMRn^{9/4}}+\cdots
\label{fvsol}\eea
where $\bal\equiv\al/(\ab\Z0\Hb)=\OMkn^{1/2}\fvn^{-1/6}\simeq3\fvn^{1/3}/(2+
\fvn)$, which is a parameter close to unity, taking values in the range
$0.973<\bal<0.999$ for solutions with $0.6<\fvn<0.9$. When $\al=0$, then
$\fv=0$ and (\ref{absol}) reduces to the standard spatially flat FLRW
solution for matter plus radiation, $\ab_{\ns{FLRW}}$. More generally we note
that as $t\to0$, then $\fv\to0$ and the series solution (\ref{absol}) differs
from the series solution for $\ab_{\ns{FLRW}}$ at the level of terms ${\rm O}
(\bal^3\OMRn^{3/2}/ \OMMn^3)\ll1$ and smaller in the coefficients of the
powers $(\Hb t)^n$, $n\ge2$.

Using (\ref{absol}), (\ref{fvsol}) series solutions for all other relevant
quantities can be obtained. For example, the void scale factor
$\av=\ab\fv^{1/3}\fvi^{-1/3}$ takes the form
\bea
\fl\frac\av{a_{\ns{vi}}}=&\frac{2}{\sqrt5}(\Hb t)+\frac{\sqrt2\,\OMMn}{15\sqrt5
\,\OMRn^{3/4}}(\Hb t)^{3/2}-\frac{209\,\OMMn^2}{5100\sqrt5\,\OMRn^{3/2}}(\Hb t)
^2\nonumber\\ \fl&
+\left(\frac{53621\sqrt2\,\OMMn^3}{3978000\sqrt5\,\OMRn^
{3/2}}+\frac{8\sqrt2\,\bal^3}{1625}\right)\frac{(\Hb t)^{5/2}}{\OMRn^{3/4}}
\nonumber\\\fl&\quad+
\left(\frac{261638807\,\OMMn^3}{28149322500\sqrt5\,\OMRn^{3/2}}+\frac{6716\,
\bal^3}{901875\sqrt5}\right)\frac{\OMMn\bal(\Hb t)^3}{\OMRn^{3/2}}+
\cdots
\label{avsol}
\eea
where $a_{\ns{vi}}\equiv\fvi^{-1/3}\abn\bal$, which to leading order is linear
in $\Hb t$ as $t\to0$, like a Milne universe, but differs at higher order.
The wall scale factor $\aw=\ab(1-\fv)^{1/3}\fwi^{-1/3}$ takes the form
\bea
\fl\frac\aw{a_{\ns{wi}}}=&\sqrt2\,\OMRn^{1/4}(\Hb t)^{1/2}+\frac{\OMMn(\Hb t)}
{3\,\OMRn^{1/2}}-\frac{7\OMMn^2(\Hb t)^{3/2}}{72\sqrt2\,\OMRn^{5/4}}\nonumber\\
\fl&+\left(\frac{5\,\OMMn^3}{216\,\OMRn^{3/2}}+\frac{4\bal^3}{75\sqrt5}
\right)\frac{(\Hb t)^2}{\OMRn^{1/2}}
-\left(\frac{91\,\OMMn^3}{6912\sqrt2\,\OMRn^{3/2}}+\frac{7
\sqrt2\,\bal^3}{225\sqrt5}\right)\frac{\OMMn(\Hb t)^{5/2}}{\OMRn^{5/4}}
\nonumber\\\fl&\quad
+\left(\frac{\OMMn^3}{243\,\OMRn^{3/2}}+\frac{11927\,\bal^3}{401625
\sqrt5}\right)\frac{\OMMn^2(\Hb t)^3}{\OMRn^2}+\cdots\label{awsol}
\eea
where $a_{\ns{wi}}=\fwi^{-1/3}\abn$, which is very close to the background
solution (\ref{absol}), differing only at ${\rm O}[(\Hb t)^2]$.

The relative expansion rate of walls and voids, and the phenomenological
lapse function, are found to be
\bea
\fl h_r=&\frac12+\frac{3\,\OMMn(\Hb t)^{1/2}}{20\sqrt2\,\OMRn^{3/4}}-\frac{463
\,\OMMn^2(\Hb t)}{6800\,\OMRn^{3/2}}+\left(\frac{332167\,\OMMn^3}{5304000\sqrt
2\,\OMRn^{3/2}}+\frac{\bal^3}{65\sqrt5}\right)\frac{(\Hb t)^{3/2}}{\OMRn^{3/4}}
\nonumber\\ \fl&
-\left(\frac{1452551123\,\OMMn^3}{50043240000\,\OMRn^{3/2}}+\frac{7329\,\bal^3}
{120250\sqrt5\,}\right)\frac{\OMMn(\Hb t)^2}{\OMRn^{3/2}}+\cdots,\label{hrsol}
\\\fl\gb=&1+\frac{2\sqrt2\,\bal^3}{5\sqrt5\,\OMRn^{3/4}}(\Hb t)^{3/2}-\frac{14
\,\OMMn\bal^3}{25\sqrt5\,\OMRn^{3/2}}(\Hb t)^2+\frac{3781\,\OMMn^2\bal^3}{5100
\sqrt{10}\,\OMRn^{9/4}}(\Hb t)^{5/2}\nonumber\\\fl&-\left(\frac{142189\,\OMMn^3
}{298350\sqrt5\,\OMRn^{3/2}}+\frac{736\,\bal^3}{8125}\right)
\frac{\bal^3(\Hb t)^3}{\OMRn^{3/2}}+\cdots
\label{gsol}
\eea

In \cite{clocks} the matter--only solution was joined to the spatially flat
FLRW solution with matter and radiation, which corresponds to the
$\bal=0$ limit above. We see that all the assumptions made in the matching
procedure were correct, with one exception. In \cite{clocks} it was assumed
that as $t\to0$ we would have $h_r\to1$, whereas it turns out that $h_r\to\frac
12$. Actually, this small difference in assumptions makes no difference to any
of the physical conclusions derived in \cite{clocks}, since the late time
solution quickly reaches a tracking limit which is largely independent of the
initial conditions. Furthermore, in deriving constraints from the early
universe, such as primordial nucleosynthesis bounds, calculations in
\cite{clocks} only required the property that $\gb\to1$ and $\fv\to0$ as $t\to
0$, which remains correct in the full solution with radiation. The exact
dependence of $h_r$ as $t\to0$ may be of relevance for determination of
features of the matter power spectrum. However, such features remain to be
determined.

It is possible to invert the series (\ref{absol}) to obtain an expression for
$\Hb t$ in terms of a series in $\ab/\abn$ at very early times. This enables
one to determine the epoch of matter--radiation equality, for example, when
$\ab_{\ns{eq}}/\abn=\OMRn/\OMMn$. It turns out that the terms in $\bal$
introduce differences of less than 0.1\% from the leading order
spatially flat FLRW result,\footnote{Here $\Hb$ is, however, the bare Hubble
constant which differs from the dressed Hubble constant, $\Hm$, according to
$\Hm=(4\fvn^2+4\fvn+4)\Hb/[2(2+\fvn)]$ \cite{sol,obs}.} $\Hb t_{\ns{eq}}=2
(2-\sqrt2)\OMRn^{3/2}/(3\OMMn^2)$.

\subsection{Numerical solutions and results}

We have determined further terms in the series expansions (\ref{absol}),
(\ref{fvsol}) and have established that they provide accurate solutions well
beyond the epoch of recombination, when compared to numerical solutions. For
practical investigations, however, it is convenient to use the series
expansions to provide initial conditions at an early time, and to integrate
the ODEs (\ref{deScale}) and (\ref{defv}) numerically. We
begin integrations after the epoch of nucleosynthesis when the universe is
radiation dominated but the number of relativistic species is no longer
affected by phase transitions, e.g., at $\Hb t\simeq5\times10^{-11}$ when the
universe is about a year old.
\begin{figure}[htb]
\vbox{\centerline{\scalebox{0.5}{\includegraphics{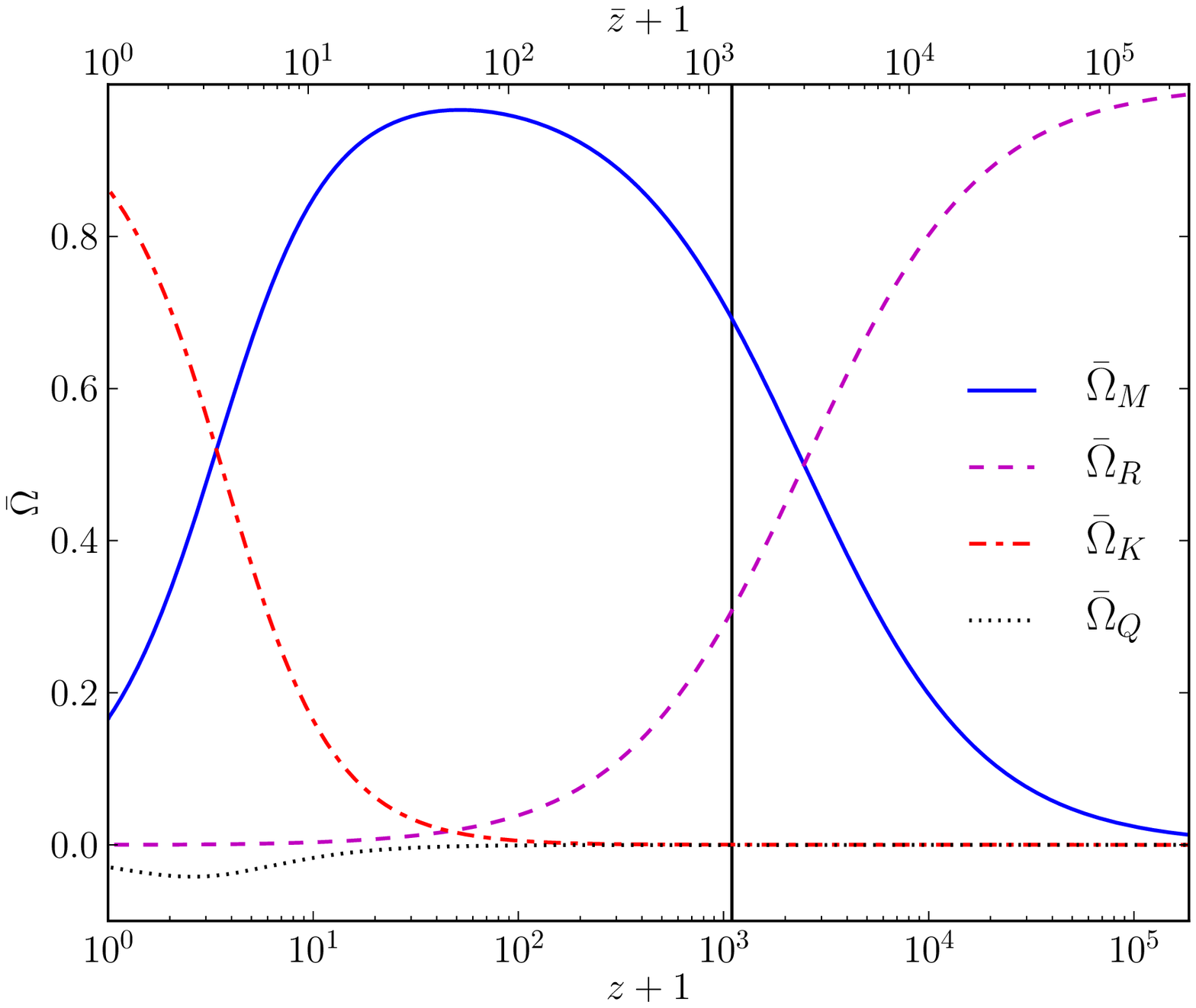}}}}
\caption{\label{fig_Om}%
{\sl Bare density parameters (\ref{om1})--(\ref{om4}) for the full numerical
solution, as a function of $z+1=\gb\abn/(\gbn \ab)$, for the dressed parameters
$\Hm=61.7\kmsMpc$, $\OmMn=0.410$. (The redshift, $z$, is the dressed parameter
measured by wall observers, and $\bar z$ is the bare redshift.)
The vertical bar at $1094.88<z<1100.46$ represents the width of the uncertainty
in the redshift of decoupling for our chosen range of $\etBg$.}}
\end{figure}

In \fref{fig_Om} we display the variation of the bare density parameters
as a function of redshift for one choice of $(\Hm,\OmMn)$ values which fit
cosmological data well. For redshifts $z\lsim10$ the bare density parameters
are essentially indistinguishable from those computed for the matter-only
solution \cite{clocks}--\cite{obs}. The full matter plus radiation solution is
certainly required to make reliable estimates of $\OMM$ and $\OMR$ at
redshifts larger than that corresponding to the maximum of $\OMM$, i.e., at
$z\gsim50$.

In earlier work we joined the matter-dominated solution \cite{sol} to
a spatially flat FLRW with matter plus radiation at the surface of last
scattering \cite{clocks,LNW,SW}. This allowed a rough estimate of the angular
diameter distance of the sound horizon, and of the effective comoving BAO scale
which is independently measured in galaxy clustering statistics. Since the
approximations were rough we simply took $z\ns{dec}\simeq1100$ as an estimate
of the epoch of photon decoupling, $\th_*\simeq0.01\,$rad as an estimate of the
angular acoustic scale, and $104\hm$ as an estimate of the effective comoving
BAO scale based on measurements for the standard cosmology known at the time
\cite{clocks,LNW}.

Given a full numerical solution including radiation, we can determine both the
epoch of photon decoupling and the subsequent baryon drag epoch directly in
parallel to our numerical integrations, using the standard physics of the
recombination era adapted to the timescape model, as discussed in \ref{recom}.
The volume--average sound horizon scale at any epoch is given by
\beq
\bD_s={\ab(t)\over\ab\Z0}{c\over\sqrt{3}}\int_0^{\mx\ns{dec}}{\dd\mx\over\mx^2
\bH\sqrt{1+0.75\,\mx\,\OMBn/\OMPn}}\,,
\eeq
where $\OM\Z{\ga0}=2g_*^{-1}\OMRn$ is the volume-average photon density
parameter at the present epoch and $g_*=3.36$ is the relative degeneracy
factor of relativistic species. We fix the value of $\OMBn=\etBg\mpr\bn\Z
{\ga0}$ in terms of the proton mass, $\mpr$, the present epoch volume--average
photon density, $\bn\Z{\ga0}$, and the baryon--to--photon ratio, $\etBg$. For
BAO measurements, the relevant comoving size of the sound horizon is that at
the baryon drag epoch, which occurs when $c\,\tad\simeq1$, where $\tad$ is the
drag depth (\ref{tad}).

\begin{figure}[htb]
\vbox{\centerline{\scalebox{0.52}{\includegraphics{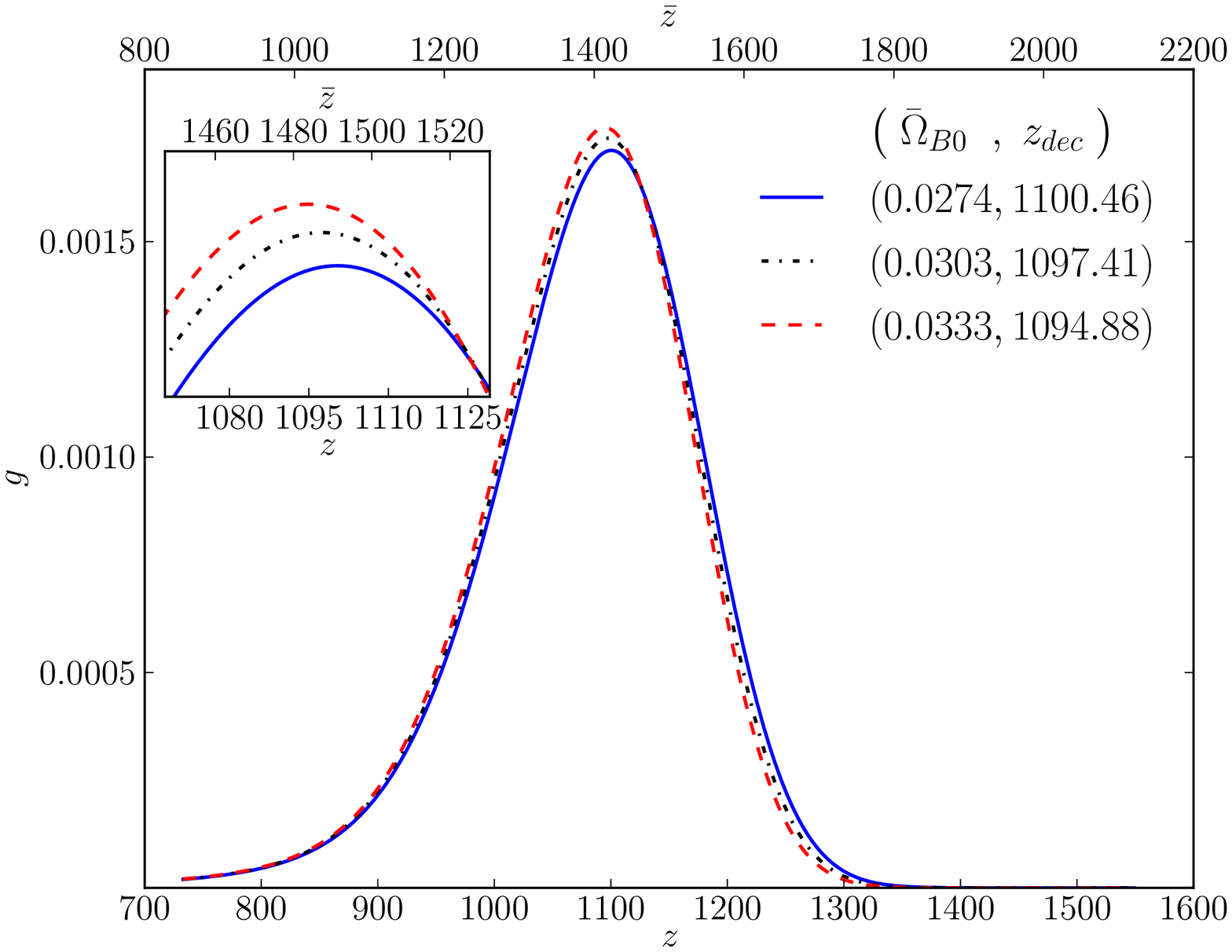}}}
\caption{\label{fig_vis}%
{\sl The visibility function (\ref{gvis}) is plotted as a function of the
dressed redshift, $z$, (and also of $\bar z$). For this example, $\Hm=61.7
\kmsMpc$ (or $\Hb=50.1\kmsMpc$), $\OmMn=0.410$ (or $\OMMn=0.167$), with three
different values of the baryon--to--photon ratio: $10^{10}\etBg=\{4.6,\,5.1,\,
5.6\}$ (which correspond to $\OMBn=\{0.0274,\,0.0303,\,0.0333\}$ respectively).
}}}\end{figure}
Examples of the estimation of the decoupling redshift, $z\ns{dec}$, from the
peak of the visibility function are shown in \fref{fig_vis} for fixed
values of ($\Hm$, $\OmMn$), with three different values of the
baryon--to--photon ratio, giving rise to different values of $\OMBn$ (or
of $\OmBn=\gbn^3\OMBn$). It is a feature of the timescape model that for a
given baryon--to--photon ratio, the baryon fraction at decoupling is increased
relative to the standard \LCDM\ model. Thus it is possible to match features
of the acoustic peaks \cite{clocks} for baryon--to--photon ratios for which
there is no primordial lithium abundance anomaly \cite{pill}. In keeping with
past work \cite{clocks,LNW}, we perform calculations for the range $\etBg=(5.1
\pm0.5)\times10^{-10}$ on the basis of constraints from light element
abundances alone\footnote{A higher value is assumed in the \LCDM\ fits of CMB
data, giving rise to the lithium abundance anomaly. While there is an intrinsic
tension in the light element data between abundances of deuterium and lithium-7
\cite{bbn2}, for the range of $\etBg$ we adopt here, all abundances fall within
2$\si$.} \cite{bbn1,bbn2}.

Using the procedures discussed in \ref{num} we have conducted numerical
integrations over the parameter space to investigate the extent to which the
timescape model parameters can be constrained using recent Planck data
\cite{Planck-parm}. We see from (\ref{ode1})--(\ref{ode3}) that since
the curvature parameter $\bal$ can be absorbed into a rescaling of the time
variable there are effectively three independent parameters, $\bH$, $\Oma$ and
$\Ora$, or equivalently $\Hb$, $\OMMn$ and $\OMRn$. Since
the bare radiation density parameter is constrained by measurements of the
CMB temperature, this leaves two independent parameters of interest. We
can take these to be either the bare parameters, $\Hb$ and $\OMMn$, or
equivalently the dressed Hubble constant, $\Hm=\gbn\Hb-\gbn^{-1}
\left.\dot\gb\right|_0$, and the dressed matter density parameter $\OmMn=\gbn^
3\OMMn$. For the figures below we will use the dressed parameters, since
$\Hm$ ideally corresponds to our measured average Hubble constant, while
$\OmMn$ is numerically closer to that of the homogeneous \LCDM\ cosmology.

\begin{figure}[htb]
\vbox{\centerline{\scalebox{0.45}{\includegraphics{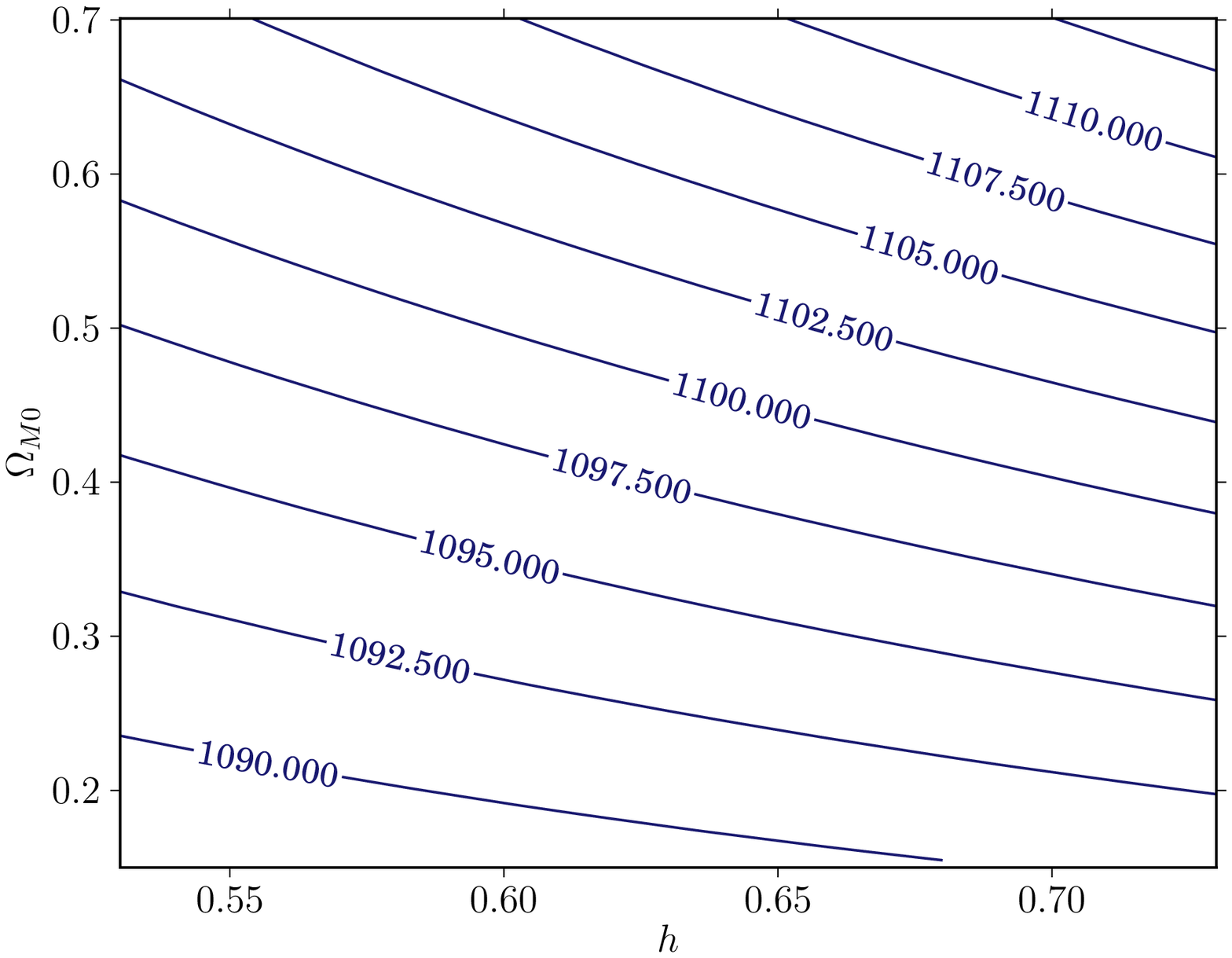}}}}
\caption{\label{dec}%
{\sl Contours of decoupling redshift, $z\ns{dec}$, in the space of
dressed parameters ($h$, $\OmMn$), (where $\Hm=100\,h\,$Mpc). Contours are
shown for the case $\etBg=5.1\times10^{-10}$.}}
\end{figure}
\begin{figure}[htb]
\vbox{\centerline{\scalebox{0.45}{\includegraphics{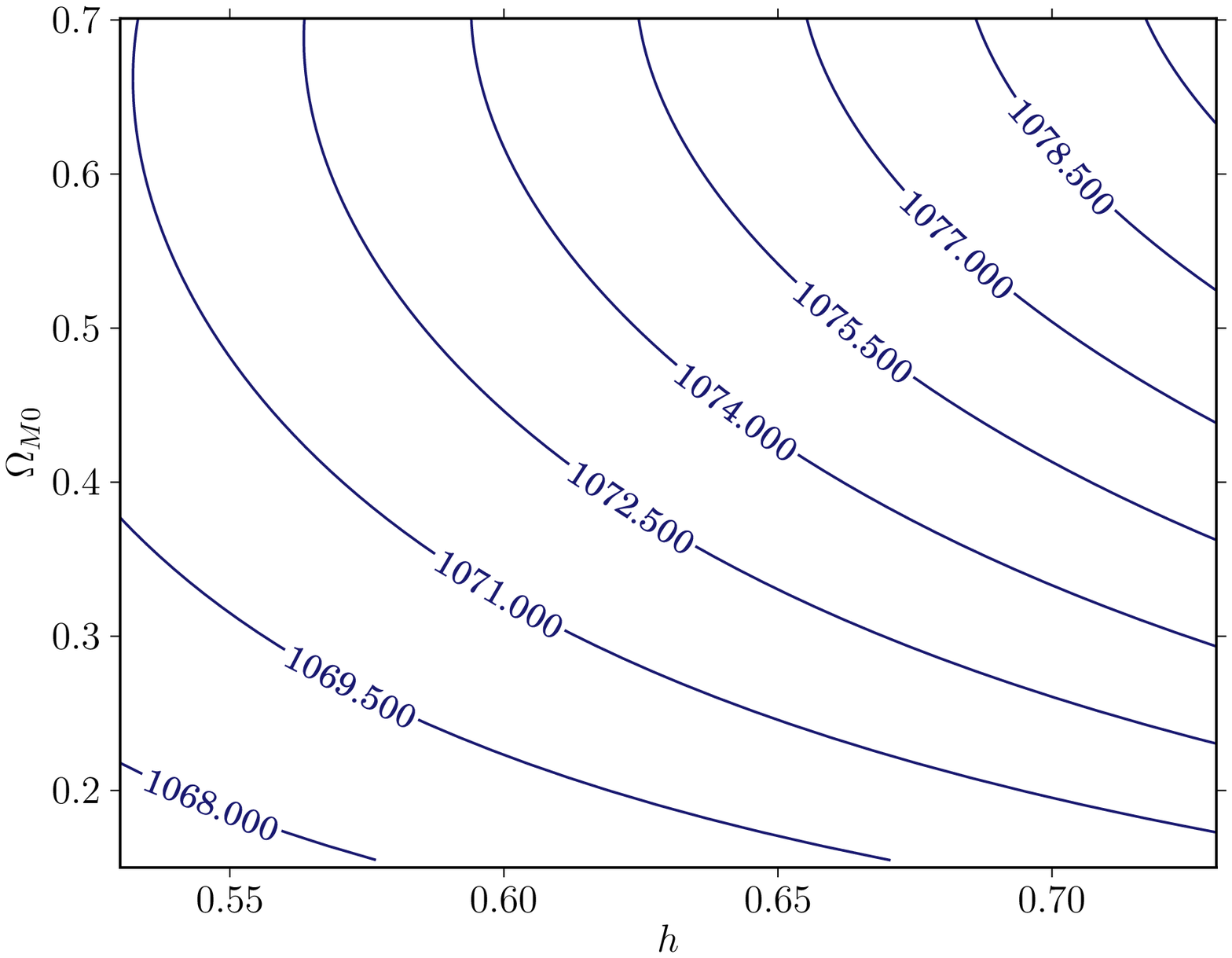}}}}
\caption{\label{drag}%
{\sl Contours of the redshift of the baryon drag epoch, $z\ns{drag}$, in the
space of dressed parameters ($h$, $\OmMn$), (where $\Hm=100\,h\,$Mpc). Contours
are shown for the case $\etBg=5.1\times10^{-10}$.}}
\end{figure}
In \fref{dec} and \fref{drag} we display contours of the decoupling
redshift, $z\ns{dec}$, and the baryon drag redshift, $z\ns{drag}$, for the
case of a fixed $\etBg=5.1\times10^{-10}$, but with the two independent
parameters ($\Hm$, $\OmMn$) varying. In each case we display the dressed
redshifts determined by wall observers such as
ourselves.

\begin{figure}[htb]
\vbox{\centerline{\scalebox{0.5}{\includegraphics{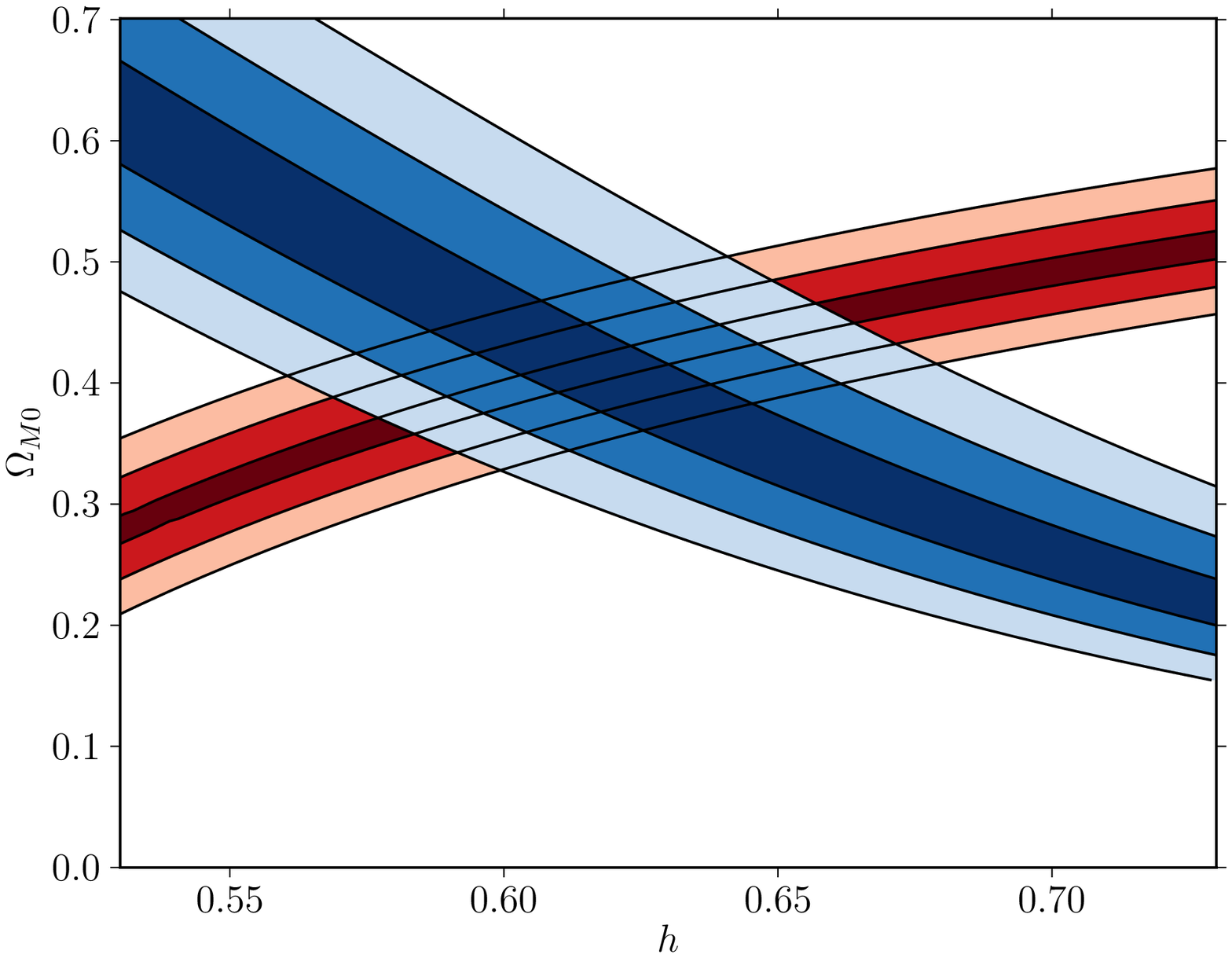}}}}
\caption{\label{cmbbao}%
{\sl Contours of ($h$, $\OmMn$) parameter values for which the angular diameter
of the sound horizon at decoupling matches the angular scale $\th_*=0.0104139$
\cite{Planck-parm} to within $\pm2$\%, $\pm4$\% and $\pm6$\% are shown in blue
(upper left to lower right). Contours of parameter values for which the
present-day effective comoving scale of the sound horizon at the baryon drag
epoch matches the value $98.88\hm$ \cite{Planck-parm} are shown in red (lower
left to upper right). In each case the baryon--to--photon ratio is assumed to
be in the range $4.6<10^{10}\etBg<5.6$.}}
\end{figure}
In \fref{cmbbao} we display two sets of contours in the ($\Hm$, $\OmMn$)
parameter space: firstly, parameters which match the acoustic scale of the
sound horizon $\th_*=0.0104139$ determined from the Planck satellite data
\cite{Planck-parm} to within $\pm2$\%, $\pm4$\% or $\pm6$\%; and secondly
parameters which similarly match the present effective comoving scale of the
sound horizon at baryon drag epoch as determined by the standard \LCDM\ model
analysis of the Planck data, namely\footnote{Since the Hubble constant $\Hm=
67.11\kmsMpc$ determined from the Planck satellite is a fit to the \LCDM\
model, any effective present comoving scale must be given in units $\hm$, as
the timescape model will generally yield a different value for $\Hm$. In all
cases, we use the values determined from the Planck data only.} $98.88\hm$
\cite{Planck-parm}. This figure updates figure 4 of \cite{SW} both in terms of
using the latest data \cite{Planck-parm}, and also in using our new full
matter--radiation solution to determine the relevant scales.

The constraints obtained from \fref{cmbbao} are not statistical constraints of
the sort that could be obtained by fully fitting the Planck data directly to
the timescape model. However, the angular scale in particular is unlikely to
differ much from that of the FLRW model analysis, and the 2\% constraint is a
reasonable estimate of uncertainty. There is greater uncertainty in the
determination of the BAO scale since the timescape model involves a potential
recalibration of the relative proportions of baryonic and nonbaryonic dark
matter which needs to be accounted for in fitting the acoustic peaks in CMB
anisotropy data\footnote{The ratio of the nonbaryonic cold dark matter
density to the baryonic density found for the \LCDM\ model with the Planck
data \cite{Planck-parm}, $\OmCn/\OmBn=5.4\pm0.2$, is well within the
range found for the timescape model in \tref{parcons}, which justifies
the approach we have taken.}.
\begin{table}
\caption{\label{parcons}Constraints on the cosmological parameters of the
timescape model obtained from a $\pm2$\% match to the angular scale,
$\th_*$, of the sound horizon at decoupling; and to a $\pm6$\% match to
the effective comoving scale, $r\ns{drag}$, of the sound horizon at the
baryon drag epoch, using recent values from the Planck satellite analysis
\cite{Planck-parm}.}
\begin{indented}
\item[]\begin{tabular}{@{}lll}
\br
Parameter &&Range\\
\mr
Dressed Hubble constant&$\Hm$&$61.7\pm3.0\,$km$/($sec$\,\cdot\,$Mpc$)$\\
Dressed matter density parameter&$\OmMn$&$0.41^{+0.06}_{-0.05}$\\
Dressed baryon density parameter&$\OmBn$&$\vp0.074^{+0.013}_{-0.011}$\\
Age of universe (galaxy/wall observer)&$\ta\ns{w0}$&$14.2\pm0.5\,$Gyr\\
Apparent acceleration onset redshift&$z\ns{acc}$&$0.46^{+0.26}_{-0.25}$\\
Present void fraction&$\fvn$&$\vp0.695^{+0.041}_{-0.051}$\\
Present phenomenological lapse function&$\gbn$&$\vp1.348^{+0.021}_{-0.025}$\\
Bare Hubble constant&$\Hb$&$50.1\pm1.7$km$/($sec$\,\cdot\,$Mpc$)$\\
Bare matter density parameter&$\OMMn$&$0.167^{+0.036}_{-0.037}$\\
Bare baryon density parameter&$\OMBn$&$\vp0.030^{+0.007}_{-0.005}$\\
Bare radiation density parameter&$\OMRn$&$\vp\left(5.00^{+0.56}_{-0.48}\right)
\times10^{-5}$\\
Bare curvature parameter&$\OMkn$&$\vp0.862^{+0.024}_{-0.032}$\\
Bare backreaction parameter&$\OM\Z{\QQ0}$&$-0.0293^{+0.0033}_{-0.0036}$\\
Nonbaryonic/baryonic matter densities ratio&$\OMCn/\OMBn$&$\vp4.6^{+2.5}_{-2.1}$\\
Age of universe (volume-average observer)&$\tn$&$17.5\pm0.6\,$Gyr\\
\br
\end{tabular}
\end{indented}
\end{table}

If we nonetheless take the 2\% constraint on $\th_*$ and the 6\% constraint on
$r\ns{drag}$ as estimates of the uncertainty, then this corresponds to the
constraints $\Hm=61.7\pm3.0\kmsMpc$, $\OmMn=0.41^{+0.05}_{-0.06}$. Constraints
on other parameters with these bounds are given in \tref{parcons}. It is
interesting to note that the constraints on ($\Hm$, $\OmMn$) are well in
agreement\footnote{For supernova analysis constraints on $\Hm$ are subject
to an overall normalization of the distance scale. In the present case,
using the normalization chosen by the authors of \cite{K09} the constraints
on both $\Hm$ and $\OmMn$ from supernovae \cite{SW} coincide with parameters
which fit the Planck data, as given in \tref{parcons}.} with the constraints
obtained from the analysis of 272 SDSS-II supernova distances \cite{K09} in the
timescape cosmology, as shown in the second panel of figure 8 of \cite{SW}. The
timescape model therefore remains competitive.

\begin{figure}[htb]
\medskip
\vbox{\centerline{\scalebox{0.5}{\includegraphics{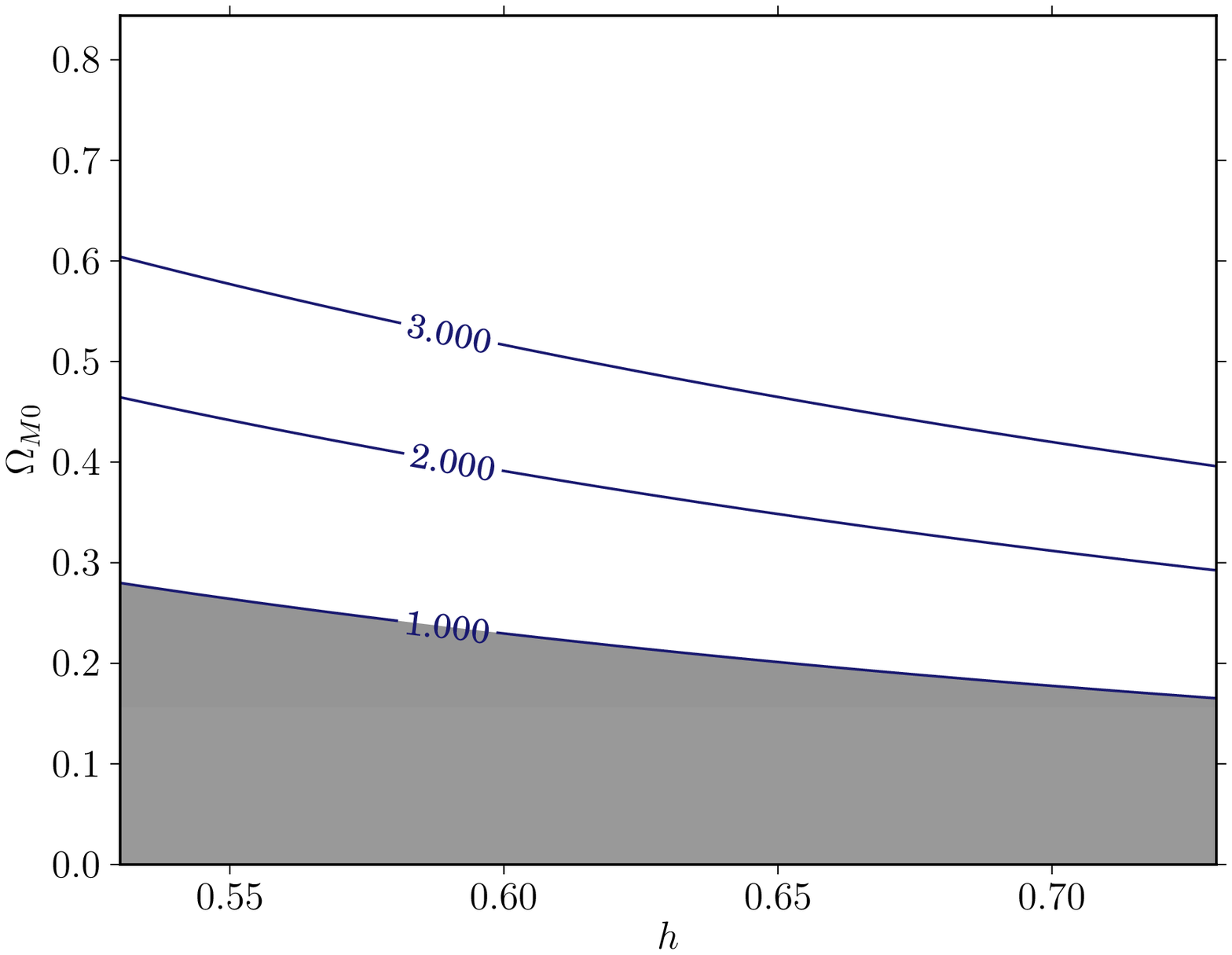}}}}
\caption{\label{ruleout}%
{\sl Contours of $\OMM/\OMR$ at $z\ns{dec}$, in the space of dressed
parameters ($h$, $\OmMn$), (where $\Hm=100\,h\,$Mpc). The shaded region with
$\OMM/\OMR<1$ is certainly ruled out.}}
\end{figure}

Even if the constraints on $r\ns{drag}$ were to change in a detailed fit
of the acoustic peaks to the timescape model, we can at least rule out
parameters by determining the ratio of matter to radiation densities at
the epoch of photon decoupling. In \fref{ruleout} we display contours
of the $\OMM/\OMR$ ratio at $z\ns{dec}$. While there is no direct
constraint on the degree to which $\OMM/\OMR$ can differ from that of the
concordance \LCDM\ cosmology, it is certainly the case that matter--radiation
equality has to occur well before decoupling in order that the standard
physics of recombination applies. In this manner we can rule out parameters
$\OmMn<0.2$ if $\Hm<65\kmsMpc$. For the parameters of \tref{parcons},
the ratio $\OMM/\OMR\simeq2$ at decoupling, which differs from the ratio
$\OMM/\OMR\simeq3$ for the concordance \LCDM\ model.

As compared to earlier best-fit values \cite{LNW}, the best-fit age
of the universe with the new constraints, $\ta\ns{w0}=14.2\pm0.5\,$Gyr, is
somewhat closer to the \LCDM\ concordance cosmology value.

\section{Conclusion}

In this paper we have derived solutions to the Buchert equations
\cite{buch1,buch2} with matter and a radiation fluid. The solutions smoothly
interpolate between a very early epoch, in which the relevant physics was that
of the standard hot big bang with an almost homogeneous FLRW background, and a
late time universe in which the average evolution is not that of a FLRW
model even though a statistical notion of homogeneity persists when one
averages on $\gsim100\hm$ scales. Both the early time series solutions
(\ref{absol}), (\ref{fvsol}) and the full numerical integrations can produce
solutions of the Buchert equations for a late-time ensemble of spatially
flat wall regions and negatively curved void regions irrespective of the
observational interpretation. Our specific calibration of parameters is that
relevant to the timescape cosmology \cite{clocks}--\cite{obs} in which the
Buchert time parameter for the statistical volume-average geometry is assumed
to differ from that of observers in bound structures, on account of
gravitational energy gradients which become significant once nonlinear
structures such as voids dominate average cosmic evolution.

The timescape cosmology is phenomenologically successful, to the extent
that it has been tested, and further tests require that its methodology is
developed in detail. This paper takes key steps towards a full analysis of
the acoustic peaks in the CMB anisotropy spectrum. The new solutions
have enabled us to directly determine the epochs of photon--electron
decoupling, $z\ns{dec}$, and of photon--baryon Compton scattering decoupling,
$z\ns{drag}$, and consequently the scale of the sound horizon at these epochs.

With these refinements the parameters of \tref{parcons} which agree
with both the acoustic scale, $\th_*$, and the BAO scale, $r\ns{drag}$, from
the Planck data \cite{Planck-parm} remain in concordance with the outcome
\cite{SW} of the fit of 277 supernovae distances from the SDSS-II survey
\cite{K09}. Furthermore, this agreement is obtained for values of the
baryon--to-photon ratio, $\etBg$, for which the primordial lithium-7
abundance is not anomalous. The timescape model therefore remains a viable
competitor to the standard cosmology.

A detailed treatment of the acoustic peaks in the CMB data may of course
still challenge the timescape cosmology, as it will certainly further
tighten the constraints. Work on this problem, which
requires a revisiting of CMB data analysis from first principles, is in
progress. As shown in \fref{ruleout}, the new results
in the present paper already allow us to rule out portions of the parameter
space, such as $\OMMn<0.2$ if $\Hm<65\kmsMpc$, which were still admissible
in previous studies \cite{LNW,SW}. Interestingly, the tightening of
constraints in the present paper has also pushed the value of the age
of the universe (in wall time) closer to that of the \LCDM\ cosmology.

It should be stressed that the value of the average Hubble constant, $\Hm=61.7
\pm3.0\kmsMpc$, inferred above\footnote{For the timescape model this
uncertainty is not yet statistical, as discussed above. Statistical
uncertainties obtained from a detailed analysis of
the Doppler peaks are likely to be of the same order as those quoted by the
Planck team for the \LCDM\ model \cite{Planck-parm}.} is a model--dependent fit
to the timescape cosmology in the same way that the value quoted by the Planck
satellite team, namely $67.4\pm1.4\kmsMpc$ \cite{Planck-parm}, is a
model--dependent fit to the FLRW cosmology. There has been much interest
about the apparent discrepancy between this \LCDM\ model value of $\Hm$
determined from the Planck satellite and the somewhat larger values from local
measurements \cite{shoes}. In the timescape model, a larger value of the
Hubble parameter is expected below the scale of statistical
homogeneity\footnote{For the parameter values of \tref{parcons} the
maximum value of the Hubble constant measured locally by a wall/galaxy
observer to the other side of a void below the statistical homogeneity
scale is $\frac32\Hb=75.2^{+2.0}_{-2.6}\kmsMpc
$.}, and this may impact the calibration of the distance scale. Such issues are
further discussed in a separate paper \cite{rest}. While no single piece of
evidence provides conclusive proof of the timescape model, a number of recent
observations which are puzzles for the standard cosmology -- the primordial
lithium-7 abundance; the local versus global values of $\Hm$; a possible
nonkinematic component to the CMB dipole \cite{rest,RS} -- are consistent with
expectations of the timescape cosmology. The results of the present paper
provide a further stepping stone to even more detailed tests of the timescape
scenario.
%------------------------------------------------
\ack
%------------------------------------------------

This work was supported by the Marsden Fund of the Royal Society of New
Zealand. We thank Teppo Mattsson, Peter Smale and Nezihe Uzun for discussions.

\appendix\section{Numerical integration}\label{num}
For the purposes of numerical integration it is convenient to write the
derivatives with respect to $x\equiv\ab/\abn$, yielding the system of
three coupled ODEs
\bea
\fl\qquad\Deriv\dd x{t'}={x\sqrt{\fFx}\over\sqrt{\fv^{1/3}x^2+\Oma x+\Ora}},
\label{ode1}\\\fl\qquad\Deriv\dd x\fv=3\Fx\sqrt{\fv\fvf},\\\fl\qquad
\Deriv\dd x\Fx={(\fFx)\left[\fv^{-1/6}\sqrt{1-\fv}+\Fx\left(\Oma+2\Ora x^{-1}
\right)\right]\over2\left(\fv^{1/3}x^2+\Oma x+\Ora\right)}-\Fx\left({3\over x}
+2x{\Fx}^2\right),\label{ode3}
\eea
in the dimensionless variables $t'\equiv\bal\Hb t$, $\fv$ and $\Fx\equiv\pt_x
\fv/[3\sqrt{\fv\fvf}]$, where $\Oma\equiv\bal^{-2}\OMMn$ and $\Ora\equiv\bal^
{-2}\OMRn$.
The wall time parameter, $\tw=\int\gb^{-1}\dd t$, may be determined also by
integrating the equation
\beq \bal\Hb\Deriv\dd x\tw={1-\fv\left(\fFx\right)\over1-\fv+x\Fx\sqrt{\fv\fvf}
}\,\Deriv\dd x{t'}\,.\label{ode4}\eeq

Given an initial estimate of $\fvn$, the tracker solution \cite{sol,obs} is
used to estimate $\OMMn\simeq4(1-\fvn)/(2+\fvn)^2$, $\OMkn\simeq9\fvn/(2+\fvn)^
2$, $\bal^2\simeq9\fvn^{2/3}/(2+\fvn)^2$ and $\Oma=4(1-\fvn)/[9\fvn^{2/3}]
$. Since $\OMRn=\kappa g_*{T_{\ns0}}^4/(\Hb{}^2\gbn^4)$, where $\kappa\equiv4
\pi^3G{\kB}^4/(45\hbar^3 c^5)$, $g_*=3.36$ and $T_{\ns0}=2.725\,$K, then given
a value of $\Hb$ and the tracker solution estimates for $\OMMn$ and $\OMkn$ we
can solve (\ref{gbn1}) to estimate $\gbn$, $\OMRn$ and $\Ora$.

Initial values of the variables are now determined at an early initial time
using the series solutions (\ref{absol}), (\ref{fvsol}), or the equivalent
series in $x$:
\bea
\fl\Hb\bal\,t=&\frac{x^2}{2\,\Ora^{1/2}}-\frac{\Oma x^3}{6\,\Ora^{3/2}}+
\frac{3\,\Oma^2x^4}{32\,\Ora^{5/2}}-\left(\frac{2\sqrt{5}}{125\,\Ora^2}+
\frac{\Oma^3}{16\,\Ora^{7/2}}\right)x^5\nonumber\\\fl&
+\left(\frac{7\sqrt{5}\,\Oma}{300\,\Ora^3}+\frac{35\,\Oma^4}{768\,\Ora^{9/2}}
\right)x^6-\left(\frac{402\sqrt{5}\,\Oma^2}{14875\,\Ora^4}+\frac{9\,\Oma^5}
{256\,\Ora^{11/2}}\right)x^{7}+\cdots\\
\fl\fv=&\frac{\sqrt{5}x^3}{25\,\Ora^{3/2}}-\frac{9\sqrt{5}\,\Oma x^4}{250\,\Ora
^{5/2}}+\frac{639\sqrt{5}\,\Oma^2x^5}{21250\,\Ora^{7/2}}-\left(\frac{6}{325\,
\Ora^3}+\frac{5553\sqrt{5}\,\Oma^3}{221000\,\Ora^{9/2}}\right)x^6\nonumber\\\fl
&+\left(\frac{11106\,\Oma}{300625\,\Ora^4}+\frac{295461027\sqrt{5}\,\Oma^4}
{13900900000\,\Ora^{11/2}}\right)x^7+\cdots
\label{msol}
\eea
We then integrate
the ODEs (\ref{ode1})--(\ref{ode4}) until the present epoch is reached at
$\xn=1$, giving the exact numerical values $t'_{\ns0}=\bal\Hb\tn$,
$\fvn$, ${\Fx}_{\ns0}$ and $\bal\Hb\tw{}_{\ns0}$. We
also have $\gbn=\left[\vphantom{\Fn^2}\right.1-\fvn+\Fn\sqrt{\fvn(1-\fvn)}\,
\left.\vphantom{\Fn^2}\right]/\left[1-
\fvn(1+{\Fn}^2)\right]$.

Only two parameters, $\Oma$ and $\Ora$, appear in the ODEs
(\ref{ode1})--(\ref{ode3}). Solutions with fixed $\Oma$, $\Ora$ therefore
represent a class of solutions which are physically equivalent under a
rescaling of the parameters $\bal$, $\OMMn$ and $\OMRn$, while keeping the
ratio $\OMMn/\OMRn$ fixed. A general solution does not have $\bH=\Hb$ at
$\xn=1$; to impose this condition we identify the right hand side
of (\ref{ode1}) at $\xn=1$ with $\bal$, from which precise values of
$\OMMn=\bal^2\Oma$, $\OMRn=\bal^2\Ora$ and $\Hb\tn$ may
be determined.

%The Buchert equations (\ref{deScale}), (\ref{defv}) possess a symmetry
%under the rescaling $t\to\alpha t$, $\OMMn\to\bal^{-2}\OMMn$, $\OMRn\to
%\bal^{-2}\OMRn$.

\section{Recombination and baryon drag epoch}\label{recom}
In \cite{clocks} the epoch of photon decoupling was set by the rough condition
consistent with the Saha equation that $z_{\ns{dec}}+1\simeq 1100$ as measured
by a wall observer, or equivalently $x_{\ns{dec}}=\zb_{\ns{dec}}+1=\gbn(1+z_
{\ns{dec}})/\gb_{\ns{dec}} $, giving $x_{\ns{dec}}\simeq 1518$ if $\gbn=1.38$,
given that $\gb_{\ns{dec}}\simeq1$.

In the present paper, we determine the epoch of photon decoupling more
precisely, using the standard physics of recombination as described, for
example, by Weinberg \cite{W08}. Since the universe is very close
to that of the standard FLRW model at this epoch there is no difference
in any physical processes, but merely in the calibration of parameters
relative to their present epoch values using the background solution.
In particular, it is convenient to work with the volume-average parameters
%$\fvn$, $\OMMn$, $\OMRn$
that appear in the Buchert equations, and therefore to work with the CMB photon
temperature, $\Tb$, as measured by a volume-average observer. In the very early
universe this is indistinguishable from the CMB photon temperature, $T$,
measured by a wall observer. However, in general the two temperatures are
related by
\beq
\Tb=\bar{\gamma}^{-1}T \,,
\eeq
giving a significant difference at the present epoch.

Following helium recombination, the ionization fraction of hydrogen is given
by $X\equiv \nb_{\ns{p}}/\nb$, where
\beq
\nb = \nb_{\ns{p}}+\nb_{\ns{H}}=0.76\,\nbB
\eeq
is the combined number density of ionized and atomic hydrogen. The relation
to the baryon number density, $\nbB$, arises from assuming that helium makes up
$24\%$ of baryons in weight. In our case, the bare baryon number density is
given by
\beq
\nbB={3\Hb^2\OMBn\over8\pi G\,m_{\ns p}}\left(\Tb\over\Tb_{\ga0}\right)^3\
\,,
\label{nv_baryon}\eeq
where $\OMBn$ is the present epoch bare baryon matter density parameter,
$\Tb_{\ga0}=\gbn^{-1}2.275\,$K and $m_{\ns p}$ the proton mass.

The ionization fraction is determined by
solving the Peebles equation \cite{Peebles_Rec}
\beq
\frac{dX(t)}{dt}=\left(\frac{\GA_{\ns{2s}}+3P\GA_{\ns{2p}}}{\GA_{\ns{2s}}+3P\GA
_{\ns{2p}}+\beta}\right)\left[-X^2+S^{-1}(1-X)\right]\Aa\,\nb\,,\label{peebles}
\eeq
where $\Aa(\Tb)$ is the effective recombination rate to the excited $2s$ and
$2p$ states; $\GA_{\ns{2s}}=8.22458\,$s$^{-1}$ and $\GA_{\ns{2p}}=4.699\times
10^8\,$s$^{-1}$ are decay rates from the $2s$ and $2p$ states;
\beq
\beta=\left(\frac{m_{\ns{e}}k_{\ns{B}}\Tb}{2\pi\hbar^2}\right)^{3/2}
\exp(-B_2/k_{\ns{B}}\Tb)\;\Aa
\eeq
is the ionization rate from the excited states;
\beq
P=\frac{8\pi \bar{H}}{3\la_{\al}^3\GA_{\ns{2p}}\nb_{\ns{1s}}}\eeq
is the photon survival probability, with $\la_{\al}=1215.682\times10^{-8}\,$cm
and $\nb_{\ns{1s}}\simeq\nb_{\ns H}$;
\beq
S\equiv 0.76 \nb_{\ns{B}}\left(\frac{m_{\ns{e}}k_{\ns{B}}\Tb}{2\pi\hbar}\right)
^{-3/2}\exp(B_1/k_{\ns{B}}\Tb)\,;
\eeq
and $B_n=m_{\ns{e}}e^4/(2\hbar^2n^2)=13.6\,n^{-2}$eV is the binding energy
of the state with principal quantum number $n$. Detailed numerical calculations
of $\Aa(\Tb)$ can be fit by the formula \cite{W08,SSS}
\beq
\Aa={1.4337\times10^{-10}\,\Tb^{-0.6166}\;\hbox{cm}^3\,\hbox{s}^{-1}\over
1+5.085\times10^{-3}\,\Tb^{0.5300}}\,,
\eeq
where $\Tb$ is given in degrees Kelvin. The Saha equation
\beq
X(1+SX)=1
\eeq
is an excellent approximation for the ionization fraction when $X$ is close to
unity during the initial equilibrium. It may therefore be used to provide at
initial value for the differential equation (\ref{peebles}) at $\Tb=4226\,$K,
or $z=1550$ after the end of helium recombination. For practical purposes,
the differential equation (\ref{peebles}) can be rewritten as an ODE in
$\Tb$ using $\dd t=-\dd\Tb/(\bH\Tb)$ or as an ODE in $x$ using (\ref{ode1})
divided by $\bal$.

The {\em visibility function},
\beq
g(t)\equiv-c\frac{d\tao}{dt}\exp(-c\tao)\;,\label{gvis}
\eeq
gives the probability that a photon last scattered at time $t\left(\Tb\right)$
when the temperature was $\Tb$,
where
\beq
\tao(t)\equiv\int_{t\left(\Tb\right)}^{t_0}\sigma_{\ns{T}}\nb_{\ns{e}}\dd t
\eeq is the {\em optical depth}, $\sigma_{\ns{T}}$ being the Thomson
scattering cross-section, and $\nb_{\ns e}=\nb_{\ns p}$ the free electron
density. The maximum of the visibility function then defines the photon
last scattering surface at $t_{\ns{dec}}$.

Acoustic fluctuations in baryons are frozen in slightly later at the baryon
drag epoch \cite{hs96} defined by the condition $c\,\tad\simeq1$, where
\beq
\tad(t)\equiv\int_{t}^{t_0}{\dot\tao\dd t\over\ab R}=
\int_{t}^{t_0}{\sigma_{\ns{T}}\nb_{\ns{e}}\dd t\over\ab R}\label{tad}
\eeq
is the {\em drag depth} and $R\equiv0.75\rh\Z B/\rh_\ga=0.75\,(\OMBn\ab)/
(\OMPn\abn)$.

\end{document}